\documentclass[showpacs,aps,amsmath,amssymb,floatfix,preprint]{revtex4}
\usepackage{graphicx} 
\begin{document}

\title{Universal Properties of Growing Networks}
\author{P. L. Krapivsky}\email{paulk@physics.bu.edu} 
\affiliation{Center for Polymer Studies and Department of Physics, 
                Boston University, Boston, MA 02215, USA}
\author{B. Derrida}
\affiliation{Laboratoire de Physique Statistique,
         Ecole Normale Sup\'erieure, 24 rue Lhomond, 75231  Paris Cedex 05, France}
         
\begin{abstract}
Networks growing according to the rule that every new node has a probability
$p_k$ of being attached to $k$ preexisting nodes, have a universal phase
diagram and exhibit power law decays of the distribution of cluster sizes in
the non-percolating phase.  The percolation transition is continuous but of
infinite order and the size of the giant component is infinitely
differentiable at the transition (though of course non-analytic). At the
transition the average cluster size (of the finite components) is
discontinuous.
\end{abstract}
\pacs{02.50.-r,  64.60.Ak,  02.10.Ox} 
\maketitle
\tighten

{\em This work is dedicated to the memory of Per Bak whose scientific work
  had an oustanding influence on the development of Statistical Physics and
  of the Theory of Complex Systems.}  
\bigskip

\section{Introduction\label{introduction}}

A number of recent works
\cite{clusters,sam,kr,3kr,bb,cb,Z} have shown that simple models of growing
networks exibit an unexpected degree of universality with a percolation
transition of infinite order.  The goal of the present paper is to summarize
and extend these recent results, and to show how the general case of a
growing network can be understood via a simple theoretical approach based on
the analysis of the differential equation (\ref{gz}) satisfied by the
generating function of distribution of cluster sizes.

Perhaps the most studied \cite{jan,ab} random network model is the random
graph introduced by Erd\"os and R\'enyi \cite{er} where each pair of 
vertices of a graph of $N$ vertices are directly connected with a
probability $c/N$. The Erd\"os and R\'enyi random graph exhibits a
percolation transition at $c=1$ and the number $NG$ of vertices in the giant
component vanishes linearly at the transition ($G \sim (c-1)$ as $c \to 1$).
Therefore it came as a surprise that the percolation transition in simple
models of {\em growing} random networks \cite{clusters} is infinitely gentle,
namely every derivative of $G$ vanishes at the transition.  Subsequent recent
works \cite{sam,kr,3kr,bb,cb,Z} confirmed that prediction for the original and
more complicated models, and additionally demonstrated another 
surprising feature: the size distribution of the connected components
has a power law decay in the non-percolating phase.  This and other
features are very reminiscent of the Berezinskii-Kosterlitz-Thouless phase
transition \cite{B,KT}; the cause of this similarity remains not understood.

The growing networks which exhibit this infinite order transition are
constructed as follows: one starts with a single node and one adds new nodes
one at each time step. When the network consists of $N$ nodes and the $N+1$st
node is added, this new node is connected to $k$ randomly chosen existing
nodes among the first $N$ nodes. Thus,  with probability $p_k$,
there are $k$ arrows emerging from the $N+1$st node, the targets of which are
$k$ nodes chosen at random among the $N$ first nodes (whether we require the
$k$ target nodes to be distinct or not has no effect on  the large $N$
properties that we discuss here).  The parameters which define the model are
the probabilities $p_k$. Natural questions one can ask
about such a network are whether it exhibits  a percolation
transition with the emergence of an infinite component and what are the critical
behaviors associated with this transition.  For a Poisson
distribution $p_k=\frac{\beta^k}{k!}\,e^{-\beta}$, the model has been
investigated in \cite{3kr,bb}, and the  case of general $p_k$'s was also recently
studied by Coulomb and Bauer \cite{cb} who obtained a number of detailed
results, in particular on the correlation between a cluster size and the
times at which the vertices of a cluster were added.

Similar models have also been studied in the mathematical literature
\cite{kw,s}, and several properties of the percolation transition have been
proved by Durrett \cite{D} and Bollob\'as {\it et al.} \cite{bjr}.  The
recent interest was  partly driven by the desire to mimic biological
networks \cite{strogatz,alon}: the Poisson network arose as the
limiting case of the protein interaction network \cite{3kr} and as a toy
model of a regulatory network \cite{bb}.

\section{The phase diagram and its critical behaviors \label{model}}

In this section, we summarize the properties of these growing networks, as
derived in sections 3-6.  The first result which emerges from the
analysis of the model is that the phase diagram as well as all the power laws
depend only on two parameters, the first two moments of the distribution
$p_k$
\begin{equation}
\beta=  \langle k \rangle = \sum_k k \  p_k  \ \ \ \ \ {\rm and} \ \ \ \ \  
\Delta =  \langle k^2 \rangle  - \langle k \rangle^2
\label{beta-Deltadef}
\end{equation}
all the other parameters being irrelevant.

Because the random variable $k$ is an integer, the variance $\Delta$ is not
only positive but it satisfies
\begin{equation}
\Delta \ge  ( \beta - [\beta]) -   ( \beta - [\beta])^2
\label{phase-dia1}
\end{equation}
where $ [\beta]$ is the integer part of $ \beta$: 
for a non-integer average $\beta$,  a
distribution concentrated on integers  has a strictly positive
variance $\Delta$.  
The distribution which minimizes $\Delta$, at fixed $\beta$, is concentrated
on the two integers $[\beta]$ and $[\beta]+1$ with $p_{[\beta]} = 1 - \beta +
[\beta]$ and $p_{[\beta]+1} = \beta - [\beta]$
leading to (\ref{phase-dia1}).

Let us summarize the main properties of the phase diagram:
\begin{enumerate}
\item {\it The non-percolating phase}
is the region
\begin{equation}
\beta  - \beta^2  \leq \Delta < \frac{1}{4}   \ \ \ {\rm for} \ \ \ \beta < \frac{1}{2}
\label{phase-dia2}
\end{equation}
In this non-percolating phase, the density $c_s$ of clusters of size $s$
decays for large $s$ like a power law
with an exponent $\tau$ which varies continuously with $\Delta$
\begin{equation}
  c_s \sim   s^{-\tau}
\ \ \ \ \ {\rm with} \ \ \ \ 
\tau = 1 + \frac{2}{1- \sqrt{1-4 \Delta}}
\label{tau}
\end{equation}

\item {\it The percolation transition line} is given by
\begin{equation}
   \Delta = {1 \over 4 }   \ \ \ {\rm and} \ \ \  \beta < {1 \over 2 }
\label{phase-dia4}
\end{equation}
along which  for large $s$
\begin{equation}
c_s \simeq {2 \over (1- 2 \beta)^2} \ {1 \over s^3 \ (\ln s)^2}
\label{cs3}
\end{equation}
\item {\it The critical boundary line} is
\begin{equation}
    \Delta = \beta - \beta^2    \ \ \ {\rm and} \ \ \   {1 \over 2 } < \beta < 1
\label{phase-dia5}
\end{equation}
along which   for large $s$,
\begin{equation}
c_s \sim s^{-1-{1 \over \beta}}
\label{cs4}
\end{equation}

\item {\it The percolating phase} covers the rest of the phase diagram, namely
\begin{equation}
\begin{cases}
\beta \leq {1 \over 2 }  
\ \ \ \ \quad  \qquad {\rm and} \ \ \  \Delta > {1 \over 4 } \cr
     {1 \over 2 }   \leq \beta < 1 
\ \ \ \  \ \ \ {\rm and} \ \ \  \Delta  > \beta - \beta^2 \cr
     1 \leq \beta \ \ \ \ \  \ \ \ \ \  \ \ \ {\rm and} \ \ \  \Delta  \ 
{\rm arbitrary} \cr
\end{cases}
\label{phase-dia3}
\end{equation}
where $c_s$ decays exponentially.  As one approaches the boundaries of the
percolating phase, the fraction $G$ of sites in the giant component vanishes
in the following ways
\begin{eqnarray}
0<\beta<{1\over 2}&  {\rm and}\quad  \Delta \to  {1 \over 4}: 
& G \sim \exp\left( - {\pi \over \sqrt{4 \Delta -1 }}\right)
\label{G1}
\\
\beta = {1\over 2} & {\rm and}\quad  \ \Delta \to {1 \over 4}:    
& G \simeq {2 e^{-1} \over \sqrt{4 \Delta -1 }}  \ 
\exp  \left(- {\pi \over 2 \sqrt{4 \Delta -1 }}\right)
\label{G2}
\\
{1\over 2}<\beta<1
& {\rm and}\quad \ \Delta\to  \beta - \beta^2\,:   &  G \sim
(\Delta - \beta + \beta^2)^{1-\beta \over 2 \beta -1} 
\label{G3}
\end{eqnarray}
\end{enumerate}

These infinite order transitions and continuously varying exponents are
reminisecent of the Berezinskii-Kosterlitz-Thouless transition \cite{B,KT}.
\begin{figure}[h]
\includegraphics[width=0.7\columnwidth]{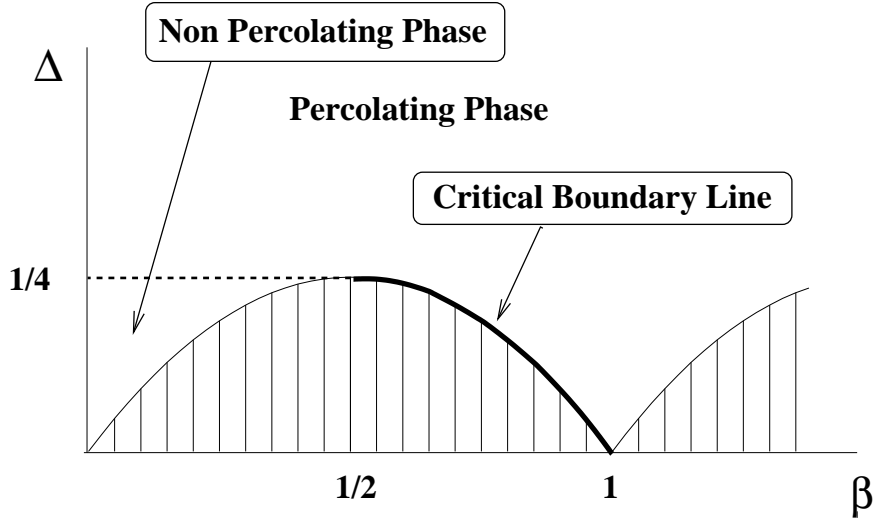}
\caption{The phase diagram}
\end{figure}

\section{The generating function of  cluster sizes}

When $N$ becomes large one can show \cite{bb} that the total number of
clusters $A_s$ of clusters of size $s$ becomes extensive ($A_s \sim N c_s$)
with fluctuations of order of order $N^{1/2}$. Therefore for large $N$, the
random variable $A_s$ is well characterized by its average $\langle
A_s\rangle$ which evolves according to
\begin{equation}
\label{As}
 \frac{d \langle A_s\rangle}{dN}=-\beta\,\frac{s\langle A_s\rangle}{N}
 +\sum_{k=0}^\infty p_k
 \sum_{s_1\ldots s_k}\prod_{j=1}^k \frac{s_j\langle A_{s_j}\rangle}{N},
\end{equation}
where the sum is taken over all $s_1\geq 1, \ldots, s_k\geq 1$ such that
$s_1+\cdots +s_k+1=s$.  On the right-hand side of (\ref{As}) the negative
term correponds to the decrease of the number of clusters of size $s$, when a
new node is introduced which connects a cluster of size $s$ to other sites,
creating that way a new cluster of size large than $s$. The positive terms
corresponds to all the ways the new node can make up a cluster of size $s$ by
connecting preexisting clusters.

Writing $\langle A_s\rangle=Nc_s$ we reduce Eqs.~(\ref{As}) to
\begin{equation}
  \label{cs}
  (1+\beta s)\,c_s=\sum_{k=0}^\infty p_k
  \sum_{s_1\ldots s_k}\prod_{j=1}^k s_j\,c_{s_j}.
\end{equation}
This allows one to determine all the $c_s$ recursively: $c_1=p_0/(1+\beta),
c_2= p_1 c_1 /(1+2 \beta), c_3 = (p_1 c_2 + p_2 c_1^2)/(1+3 \beta)$, etc...
The infinite set of Eqs.~(\ref{cs}) can be converted into a single
differential equation for the generating function
\begin{equation}
\label{gzdef}
  g(z)=\sum_{s=1}^\infty sc_s\,e^{sz}.
\end{equation}
Multiplying both sides of Eq.~(\ref{cs}) by $s\,e^{sz}$ and summing over $s$
gives
\begin{equation}
\label{gz}
  g+\beta g'=e^z\left(\mathcal{ P}[g] +g'\,\mathcal{ Q}[g]\right),
\end{equation}
where $g'=dg/dz$, the function $\mathcal{ P}[g]$ is the generating function
of the $p_k$'s and $\mathcal{ Q}[g]$ its derivative:
\begin{equation}
\label{PQ}
\mathcal{ P}[g]=\sum_{k\geq 0} p_k\,g^k, \qquad 
\mathcal{ Q}[g]= \frac{d \mathcal{ P}[g]}{dg}\,.
\end{equation}
  
Up to a change of notation, Eq.~(\ref{gz}) is the same as Eq.~(17) of
Ref.~\cite{cb}.  We will see below that all the properties summarized in
section 2 follow from (\ref{gz}).

\section{The average cluster size}

If  a node is chosen at random (ouside the giant component) the
 average size $\langle s\rangle$ of the cluster it belongs to is given by
\begin{equation}
\label{savdef}
\langle s\rangle=\frac{\sum s^2 c_s}{\sum s c_s}=\frac{g'(0)}{g(0)}
\end{equation}

The region without the giant component is characterized mathematically by
$g(0)=1$. If we replace $g(z)$ by its expansion
$g(z)= 1 + g'(0) z + o(z)$ 
in  Eq.~(\ref{gz}),
and use that
\begin{equation}
\label{Pg}
\mathcal{ P}[g]=1+\beta(g-1)+\frac{1}{2}\, \gamma (g-1)^2+\ldots
\end{equation}
with \begin{equation}
\gamma= \Delta + \beta^2 - \beta \label{gamma-def}
\end{equation}
 we obtain a quadratic equation satisfied by $g'(0)$ 
\begin{equation}
\label{g1}
(\Delta + \beta^2 - \beta) [g'(0)]^2+(2\beta-1)g'(0)+1=0.
\end{equation}
Equation (\ref{g1}) has no positive solution when
$\beta\geq \frac{1}{2}$ [see (\ref{phase-dia1})].  For $\beta\leq \frac{1}{2}$ and 
$\Delta < \frac{1}{4}$, the solution to  Eq.~(\ref{g1}) reads
\begin{equation}
\label{sav}
\langle s \rangle = g'(0)=\frac{1-2\beta-\sqrt{1-4\Delta}}{2 (\Delta +
\beta^2 - \beta)}= {2 \over 1 - 2 \beta + \sqrt{1-4\Delta}} \,. 
\end{equation}
This  suggests that different behaviors occur below, at, and above
the critical line (\ref{phase-dia4}). We shall see  in section 6
   that the giant component is born when the
critical line (\ref{phase-dia4}) is crossed.
When the line (\ref{phase-dia4})  is approached from below, i.e. $ \Delta
\to  \frac{1}{4} -0$,
one sees that
\begin{equation}
 \langle s \rangle \to {2 \over 1 - 2 \beta}
\label{sa1}
\end{equation}

Turn now to the phase with the giant component. Setting $z=0$, 
Eq.~(\ref{gz}) gives  $g'(0)$ in terms of  the size $G=1-g(0)$ of the
giant component 
\begin{equation}
\label{sav1}
g'(0)=\frac{\mathcal{ P}[1-G]+G-1}{\beta-\mathcal{ Q}[1-G]}.
\end{equation}
There is no explicit formula for the average size in the percolating phase
due to the lack of explicit expression for $G$. However, near the critical
line (\ref{phase-dia4}) one can use the fact that $G\to 0$ to simplify
Eq.~(\ref{sav1}) and get as $\Delta \to \frac{1}{4} +0$
\begin{equation}
\langle s\rangle\to \frac{1-\beta}{\Delta + \beta^2 - \beta}
=\frac{2}{1-2\beta}\,\frac{2-2\beta}{1-2\beta}
\label{sa2}
\end{equation}
Comparing  (\ref{sa1},\ref{sa2}) shows that the average size $\langle s
\rangle$ of finite components jumps
discontinuously as one crosses the percolation transition line
(\ref{phase-dia4}).

\section{Component Size Distribution}

We now examine the component size distribution $c_s$.
  
\subsection{Sub-critical regime $\beta<\frac{1}{2} $ and $\Delta < {1 \over 4}$ }

The large $s$ behavior of $c_s$ can be read off from the behavior of the
generating function (\ref{gzdef}) in the $z\to 0$ limit.  A power law decay
(see Appendix B)
\begin{equation}
\label{asym}
c_s\sim B s^{-\tau}\quad{\rm as} \quad s\to\infty 
\end{equation}
implies the following small $z$ expansion of the generating function
(\ref{gzdef}):
\begin{eqnarray}
\label{expan} 
g(z)=1+g'(0)z+B \ \Gamma(2-\tau)(-z)^{\tau-2}+\ldots.
\end{eqnarray}
Substituting this expansion into Eq.~(\ref{gz}) or rather into
\begin{equation}
\label{gz1}
g- e^z \mathcal{ P}[g] =g'\,\left( e^z \mathcal{ Q}[g] - \beta \right)
\end{equation}
we find by balancing the contributions of the order of $(-z)^{\tau-2}$

\begin{equation}
\label{tau1}
\tau-2 =- \frac{\beta -1 + (\Delta + \beta^2 - \beta) g'(0)}
{\beta  + (\Delta + \beta^2 - \beta)g'(0)} 
\end{equation}
which using (\ref{sav})  gives  (\ref{tau}). 

Thus in contrast to ordinary critical phenomena, one gets a power-law
(\ref{asym}) in the whole non-percolating phase and the  exponent $\tau$
given in (\ref{tau})  depends continuously  on the single parameter
$\Delta$.

The amplitude $B$ in  (\ref{asym}) is, as usual for critical amplitudes,
more difficult to determine. It can  nevertheless be  calculated
\cite{cb}  along the line
\begin{equation}
\label{line}
\Delta = \beta - \beta^2 \ \ \ {\rm and} \ \ \ \beta < {1 \over 2}
\end{equation}
where only $p_0$ and $p_1$ are non-zero.  
Then
the recursion (\ref{cs}) reduces to \\ \makebox{
$(1+ \beta s) c_s = \beta (s-1) c_{s-1}$} which is easily solved 
\begin{equation}
\label{cs-sol}
c_s =\frac{1-\beta}{\beta}\,{\Gamma(1+ {1\over \beta}) \; 
\Gamma(s) \over \Gamma(s+ 1+ {1\over \beta}) }
 \  \ \  \simeq 
 \ \ \    {1- \beta \over \beta} {\Gamma(1+ {1\over \beta}) \over s^{ 1+ {1\over \beta}} }
\  \ \ \ {\rm for \ large \ } s
\end{equation}

As claimed in (\ref{cs4}), expressions (\ref{cs-sol}) remain valid along the
critical boundary line (\ref{phase-dia5}), where only two probabilities $p_0$
and $p_1$ are non-zero.  Note (see Appendix B) that for small $z$ and $\beta
> {1 \over 2}$, one has
\begin{eqnarray}
g(z) &=& 1 + {1 - \beta \over \beta}
 \Gamma\left(1 + {1 \over \beta}\right) 
 \Gamma\left(1 - {1 \over \beta}\right) 
(-z)^{{1 \over \beta} -1}  +... 
\label{gz2}
\end{eqnarray}
At $\beta={1 \over 2}$, equation (\ref{cs-sol}) becomes $c_s=2/[s(s+1)(s+2)]$
leading to \\ \makebox{$g= 2 e^{-z} - 1 - 2 e^{-z} ( 1 - e^{-z}) \log (e^{-z}
  -1) $ } which gives for small negative $z$
\begin{equation}
g= 1 - 2 z \log(-z) - 2z + ...
\label{gz3}
\end{equation}
 \subsection{Critical line $0<\beta<\frac{1}{2}$ and  $\Delta= {1 \over 4} $} 
We write as in Appendix A
\begin{equation}
\label{gh} 
g(z)=1+ z v(z)
\end{equation}
and we obtain (\ref{form3}) an implicit form of $v(z)$
\begin{equation}
\ln( 1 - (1/2 - \beta) v) +\ln(-z)
+\frac{1}{( 1 - 2 \beta)[ 1 - (1/2 - \beta) v ]}
= A(\beta)
\label{vsol}
\end{equation}
The integration constant $A(\beta)$ cannot be determined without integrating the full
equation (\ref{gz}) with appropriate boundary condition: $g\to c_1\,e^z$ with
$c_1=\frac{p_0}{1+\beta}$ as $z\to -\infty$.  

{}From (\ref{gh},\ref{vsol}), we get the small $z$ of $g(z)$:
\begin{eqnarray}
\label{exp1} 
g(z)=1+\frac{2}{1-2\beta}\,z+\frac{2}{(1-2\beta)^2}\,\frac{z}{\ln(-z)}+\ldots. 
\end{eqnarray}
Inverting this expansion  (see Appendix B) yields (\ref{cs3}).

Thus, the component size distribution acquires a logarithmic correction in
the critical regime with a remarkable degree of universality: it depends only
on the average connectivity $\beta$ of the network.
  
\subsection{Super-critical regime $0<\beta<\frac{1}{2}$ and $\Delta > {1 \over 4}$}
 
Above the phase transition point, both $g(0)=1-G$ and $g'(0)$ are finite.
Repeatedly differentiating Eq.~(\ref{gz}) and setting $z=0$ we find that all
following derivatives are finite as well.  This implies that for $\Delta >
1/4$, the component size distribution decays faster than any power law.  We
now argue that
\begin{equation}
\label{asym2}
c_s \sim s^{-5/2}  e^{-s/s_*}\quad{\rm as} \quad s\to\infty.
\end{equation}

An exponential factor in the component size distribution, $c_s \propto
e^{-s/s_*}$, shifts a singularity of the generating function $g(z)$ to
$z_*=1/s_*$.  {}From Eq.~(\ref{gz}) which is useful to re-write in the
form
\begin{equation}
\label{gzC} 
g'=\frac{e^z \mathcal{ P}[g] -g}{\beta- e^z \mathcal{ Q}[g]}
\end{equation}
we see that the singularity arises when the denominator on the right-hand
side of (\ref{gzC}) vanishes: $\beta=e^{z_*}\,\mathcal{ Q}[g_*]$ where
$g_*=g(z_*)$.  The derivative on the left-hand side of (\ref{gzC}) should be
therefore singular suggesting an algebraic asymptotic $g(z)-g_*\propto
(z_*-z)^\alpha$ with $\alpha<1$. Plugging this into (\ref{gzC}) gives
$\alpha=1/2$. This type of singularity corresponds to the $s^{-3/2}$ decay of
the sequence $sc_s e^{sz_*}$ (Appendix B) and hence we finally obtain
(\ref{asym2}).  A more detailed analysis would allow to see that $s_* \sim
1/G$ diverges as the percolation line is approached \cite{cb}.

\section{Giant Component}

To determine the size of the giant component $G$ we need to understand the
behavior of solutions to Eq.~(\ref{gz}) near $z=0$ and this can be done
analytically near the transition line or near the critical boundary line (see
Appendix A).

\subsection{ For $0<\beta<\frac{1}{2}$ and  $\Delta >
{1 \over 4}$ }
If we write $g(z)$ as (\ref{gh}), set $\epsilon =\Delta -1/4 \ll 1$ 
and we use expression (\ref{form2}) for $v(z)$, we get 
\begin{equation}
\ln( 1 - (1/2 - \beta) v) +\ln(-z) - { \pi \over  2 \sqrt{\epsilon}}
+\frac{1}{( 1 - 2 \beta)[ 1 - (1/2 - \beta) v ]} 
= \ln\left[G \,\sqrt{\frac{1}{2} - \beta} \,\right] 
\label{vsol1}
\end{equation}
For this solution to be consistent with (\ref{vsol}) in the limit $\epsilon
\to 0$ the size $G$ of the giant component should satisfy
\begin{equation}
\label{giant1}
G\sim \frac{2\,e^{A(\beta)}}{1-2\beta}\,\, 
\exp\left\{-\frac{\pi}{2\sqrt{\epsilon}}\right\}
= \frac{2\,e^{A(\beta)}}{1-2\beta}\,\, 
\exp\left\{-\frac{\pi}{\sqrt{4 \Delta -1}}\right\}
\end{equation}
as claimed in (\ref{G1}).  Therefore the transition is of infinite order
since all derivatives of $G$ vanish as $\Delta \to 1/4$.  The behavior
(\ref{giant1}) appears quite universal as it was observed numerically
\cite{clusters} and confirmed analytically \cite{sam,kr,3kr,bb,cb} for other
growing networks.

\subsection{$\beta=\frac{1}{2}$ and $\Delta >{1 \over 4}$} 

In this region we may use either (\ref{form1bis}) or (\ref{form2}) to obtain
for small $\gamma$

\begin{equation}
 \ln(-z) - {\pi \over 4 \sqrt{\gamma}} + {v \over 2}  = \ln(\sqrt{\gamma} G )
\end{equation}

To be consistent with (\ref{gz3}), i.e. with $v=-2-2\ln(-z)$ in the limit
$\gamma= \Delta- 1/4\to 0$, the size of the giant component should statisfy
\begin{equation}
\label{giant2}
G(\gamma)\simeq \frac{e^{-1}}{\sqrt{\gamma}}\, 
\exp\left[-\frac{\pi}{4\sqrt{\gamma}}\right]
\end{equation}
as claimed in (\ref{G2}).  In comparison with (\ref{giant1}) there is a
factor 1/2 in the exponential and the prefactor is determined here
analytically.  The transition still is of infinite order.

\subsection{$\frac{1}{2}<\beta<1$ and $\Delta > \beta-\beta^2$} 

In this range of parameters one can use (\ref{form1}). When one approaches
the critical boundary line (\ref{phase-dia5}), i.e. for small $\gamma$,
expression (\ref{form1}) becomes
\begin{equation}
\label{vb-sol}
\frac{\beta}{2\beta-1}\,\ln[1+(2\beta-1)v] +\ln(-z)
=\ln\left\{\gamma^{-\frac{1-\beta}{2\beta-1}}\,G\right\}
+\frac{\ln(2\beta-1)}{2\beta-1}
\end{equation}
To be consistent with  (\ref{gz2})
\begin{equation}
\label{vb-hyper}
v\sim C(-z)^{\frac{1}{\beta}-2},\quad 
C=
 -{(1 - \beta )
 \Gamma\left(1 + {1 \over \beta}\right)
 \Gamma\left(1 - {1 \over \beta}\right)
\over \beta}
 =- {1 - \beta \over \beta^2}
 {\pi \over \sin( {\pi \over \beta} )}
\end{equation}
  for small $z$,
one needs  the size $G$ of the giant component 
to scale as
\begin{equation}
\label{giant3}
G(\beta,\gamma)\sim D\,\gamma^\nu, \qquad \nu=\frac{1-\beta}{2\beta-1}\,, 
\end{equation}
where $D=C^{\frac{\beta}{2\beta-1}}\,(2\beta-1)^{-\frac{1-\beta}{2\beta-1}}$
and $C$ is given by (\ref{vb-hyper}).  The exponent $\nu$ is again universal
as it depends only on the average connectivity $\beta$ but not on any other
parameter of the distribution $p_k$. Yet it varies continuously with $\beta$
along the critical boundary line (\ref{phase-dia5}) thereby showing a richer
behavior than in random graph models where $\nu=1$.

\appendix
\label{A}
\section{Analysis of (\ref{gz}) for $z$ small and $1-g$ small }
In this appendix we analyze the solution of (\ref{gz}) when both
$z$ and  $1-g$ are small.
If we write
\begin{equation}
g(z) = 1 + z v(z) \;,
\label{vdef}
\end{equation}
equation (\ref{gz}) gives to leading order
$$(1 - v + 2 \beta v + \gamma v^2)  z + (\beta + \gamma v ) v' z^2=0 $$
so that $v(z)$ satisfies
\begin{equation}
\label{vg1}
\frac{\gamma v+\beta}{\gamma v^2+(2\beta-1)v+1}\,\,dv
+\frac{dz}{z}=0\,.
\end{equation}
If the size $G$ of the giant component is non-zero, $\lim_{z\to 0}zv=g(0)-1=-G$
and this fixes the constant of integration  and leads to the following expression  
 valid for $\beta > 1/2$, $\gamma$ small and $\gamma < (\beta -1/2)^2$:
\begin{eqnarray}
&&\frac{1}{2}\,\ln[\gamma v^2+(2\beta-1)v+1] +\ln(-z)
\label{form1} \\
&&-\frac{1}{2\sqrt{(2\beta-1)^2-4\gamma}}\,\ln
\frac{ 2\beta-1 + 2 \gamma v  + \sqrt{(2\beta-1)^2-4\gamma}}
{ 2\beta-1 + 2 \gamma v -\sqrt{(2\beta-1)^2-4\gamma}}
=\ln(\sqrt{\gamma}\,G).
\nonumber
\end{eqnarray}
If one tries to analytic continue (\ref{form1}) to the region $\beta >1/2$,
$\gamma $ small and $\gamma > (\beta -1/2)^2$ one gets
\begin{eqnarray}
&&\frac{1}{2}\,\ln[\gamma v^2+(2\beta-1)v+1] +\ln(-z) 
\label{form1bis} \\
&&+\frac{i}{2\sqrt{4 \gamma -(2\beta-1)^2}}\,\ln
\frac{ 2 \beta -1 + 2\gamma v + i \sqrt{4 \gamma -(2\beta-1)^2}}
{ 2 \beta -1 + 2\gamma v -i \sqrt{4 \gamma -(2\beta-1)^2}}
=\ln(\sqrt{\gamma}\,G).
\nonumber
\end{eqnarray}
and to the region $\beta <1/2$, $\gamma $ small and $\gamma > (\beta -1/2)^2$
one gets (the extra term coming from the continuation of the logarithm)
\begin{eqnarray}
&&\frac{1}{2}\,\ln[\gamma v^2-(1-2\beta)v+1] 
+\ln(-z) - \frac{\pi}{\sqrt{4 \gamma -(1-2\beta)^2}}\
\label{form2} \\
&&+\frac{i}{2\sqrt{4 \gamma -(1-2\beta)^2}}\,\ln
\frac{1 - 2 \beta - 2\gamma v - i \sqrt{4 \gamma -(1-2\beta)^2}}
{1 - 2 \beta - 2\gamma v +i \sqrt{4 \gamma -(1-2\beta)^2}}
=\ln(\sqrt{\gamma}\,G).
\nonumber
\end{eqnarray}
Lastly when $G=0$ and $\gamma=(\beta -1/2)^2 $ 
one gets by integrating (\ref{vg1})
\begin{eqnarray}
\ln( 1 - (1/2 - \beta) v) +\ln(-z) 
+\frac{1}{( 1 - 2 \beta)[ 1 - (1/2 - \beta) v ]}
= A(\beta)
\label{form3} 
\end{eqnarray} 
where $A(\beta)$ is an integration constant.

\section{Extracting the Asymptotics}
\label{B}

Consider the generating function 
\begin{equation}
\label{Azdef}
A(z)=\sum_{s=1}^\infty a_s\,e^{sz}.
\end{equation} 
The dominant singularity of the generating function allows one  to extract the
large $s$ asymptotic of $a_s$. This can be done by a variety of techniques
 \cite{Hardy,FO}. Here we give an elementary exposition that is sufficient to
extract the asymptotics used in this paper. 

Let us first see what kind of the singular behavior is associated with the
power-law asymptotic
\begin{equation}
\label{As-large}
a_s\simeq A\,s^{\alpha}\qquad {\rm as} \quad s\to\infty.
\end{equation} 
If $\alpha>-1$, then $A(z)$ diverges as $z\uparrow 0$. The dominant
contribution is found by replacing summation by integration:
\begin{equation}
\label{a>-1}
A(z)\to \int_{0}^\infty ds\, A\,s^{\alpha}\,e^{sz}
=A\,\Gamma(1+\alpha)\,(-z)^{-1-\alpha}\,.
\end{equation} 
If $ -2<\alpha < -1$, the sum $\sum_{s\geq 1} a_s$ converges and instead of
(\ref{a>-1}) we get
\begin{equation}
\label{a>-2}
A(z)=A(0)+A\,\Gamma(1+\alpha)\,(-z)^{-1-\alpha}+\ldots.
\end{equation} 
Similarly for $-3 < \alpha < -2$,
\begin{equation}
\label{a>-3}
A(z)=A(0)+A'(0)z+A\,\Gamma(1+\alpha)\,(-z)^{-1-\alpha}+\ldots
\end{equation} 
where of course $A(0)=\sum_{s\geq 1} a_s$ and $A'(0)=\sum_{s\geq 1} sa_s$.
Thus if the dominant {\em singular} term has the power-law form
$(-z)^{-1-\alpha}\,A\,\Gamma(1+\alpha)$, the asymptotic must have the
form
(\ref{As-large}).

Above we assumed that $\alpha$ is not a negative integer. Otherwise
logarithms can arise. For instance, if $\alpha=-1$ we get
\begin{equation}
\label{a=-1}
A(z)\to \int_{1}^\infty ds\, A\,s^{-1}\,e^{sz}
\to -A\,\ln(-z)\,.
\end{equation} 
Imagine now that the dominant singular term is a power of $\ln(-z)$. 
It is tempting to test the asymptotic 
\begin{equation}
\label{As-large-a}
a_s\simeq A\,s^{-1}(\ln s)^{-a}\qquad {\rm as} \quad s\to\infty.
\end{equation} 
For $a\leq 1$, it leads to 
\begin{equation}
\label{a<}
A(z)\to A\int^{1/(-z)} \frac{ds}{s\,(\ln s)^a}
\to A\times
\begin{cases}
(1-a)^{-1}[-\ln(-z)]^{1-a} & a<1;\cr
\ln[-\ln(-z)]& a=1.
\end{cases}
\end{equation} 
For $a>1$, the sum $\sum_{s\geq 1} a_s$ converges and we get
\begin{equation}
\label{a>1}
A(z)=A(0)-\frac{A}{a-1}\,[-\ln(-z)]^{1-a}+\ldots.
\end{equation} 

As an example (\ref{exp1}), imagine that we have found the small $z$
expansion of the generating function $g(z)=\sum s\,c_s\,e^{sz}$ that reads
\begin{equation}
\label{example}
g(z)=A_0+A_1\,z+A\,\frac{z}{\ln(-z)}+\ldots.
\end{equation} 
Diffirentiating gives
\begin{equation}
\label{diff-example}
g'(z)=A_1+A\,\frac{1}{\ln(-z)}+\ldots
\end{equation} 
which, in conjuction with (\ref{a>1}), shows that $a=2$. Using the asymptotic
(\ref{As-large-a}) and $g'(z)=\sum s^2\,c_s\,e^{sz}$ we get as in
(\ref{cs3})
\begin{equation}
\label{As-example}
c_s\simeq A\,s^{-3}(\ln s)^{-2}\qquad {\rm as} \quad s\to\infty.
\end{equation}

\end{document}